\documentclass[useAMS,usenatbib,usegraphicx]{mn2e}

\newcommand{\esn}{$E_{51}$}

\newcommand{\msun}{${\rm M}_\odot$}
\newcommand{\msunyr}{${\rm M}_\odot\ {\rm yr}^{-1}$}
\newcommand{\be}{\begin{equation}}
\newcommand{\ee}{\end{equation}}
\newcommand{\circa}{$\sim$}
\newcommand{\cmt}{cm$^{-3}$}
\newcommand{\kms}{km s$^{-1}$}
\newcommand{\cl}{_{\rm cl}}
\renewcommand{\th}{$T_{\rm h}$}       \newcommand{\Th}{T_{\rm h}}
       \newcommand{\Eh}{E_{\rm hot}}
\newcommand{\tc}{$T_{\rm c}$}         \newcommand{\Tc}{T_{\rm c}}
\newcommand{\nh}{$n_{\rm h}$}         \newcommand{\Nh}{n_{\rm h}}
\newcommand{\nc}{$n_{\rm c}$}         \newcommand{\Nc}{n_{\rm c}}
\newcommand{\tff}{$t_{\rm dyn}$}      \newcommand{\Tff}{t_{\rm dyn}}
\newcommand{\fcool}{$f_{\rm cool}$}   \newcommand{\Fcool}{f_{\rm cool}}
\newcommand{\fstar}{$f_\star$}        \newcommand{\Fstar}{f_\star}
\newcommand{\tinf}{$t_{\rm inf}$}     \newcommand{\Tinf}{t_{\rm inf}}
\newcommand{\tcool}{$t_{\rm cool}$}   \newcommand{\Tcool}{t_{\rm cool}}
     \newcommand{\Mhot}{M_{\rm hot}}
   \newcommand{\Mcold}{M_{\rm cold}}
        
   \newcommand{\Mhalo}{M_{\rm halo}}
\newcommand{\tleak}{$t_{\rm leak}$}   \newcommand{\Tleak}{t_{\rm leak}}
\newcommand{\frest}{$f_{\rm rest}$}   \newcommand{\Frest}{f_{\rm rest}}
\newcommand{\fcoll}{$f_{\rm coll}$}   \newcommand{\Fcoll}{f_{\rm coll}}
\newcommand{\fevap}{$f_{\rm evap}$}   \newcommand{\Fevap}{f_{\rm evap}}
\newcommand{\krh}{$k_{\rm rh}$}       \newcommand{\Krh}{k_{\rm rh}}
\newcommand{\kres}{$k_{\rm resv}$}    \newcommand{\Kres}{k_{\rm resv}}
\newcommand{\ktr}{$k_{\rm trigger}$}  \newcommand{\Ktr}{k_{\rm trigger}}
\newcommand{\poro}{$Q_{\rm sb}$}      
     \newcommand{\Fion}{f_{\rm ion}}
\newcommand{\dmhot}{$\dot{M}_{\rm hot}$}
\newcommand{\Dmhot}{\dot{M}_{\rm hot}}
\newcommand{\dmcold}{$\dot{M}_{\rm cold}$} 
\newcommand{\Dmcold}{\dot{M}_{\rm cold}}

\newcommand{\Dmstar}{\dot{M}_\star}

\newcommand{\Dmhalo}{\dot{M}_{\rm halo}}      

\newcommand{\Dmcool}{\dot{M}_{\rm cool}}
\newcommand{\dmsnpl}{$\dot{M}_{\rm snpl}$}

\newcommand{\Dmev}{\dot{M}_{\rm evap}}

\newcommand{\Dmrest}{\dot{M}_{\rm rest}}
\newcommand{\dmsfr}{$\dot{M}_{\rm sf}$}
\newcommand{\Dmsfr}{\dot{M}_{\rm sf}}
\newcommand{\dminf}{$\dot{M}_{\rm inf}$}
\newcommand{\Dminf}{\dot{M}_{\rm inf}}
\newcommand{\dmleak}{$\dot{M}_{\rm leak}$}
\newcommand{\Dmleak}{\dot{M}_{\rm leak}}
\newcommand{\desn}{$\dot{E}_{\rm sn}$}
\newcommand{\Desn}{\dot{E}_{\rm sn}}

\newcommand{\Decool}{\dot{E}_{\rm cool}}
\newcommand{\desnpl}{$\dot{E}_{\rm snpl}$}

\newcommand{\Dehot}{\dot{E}_{\rm hot}}

\newcommand{\Deleak} {\dot{E}_{\rm leak}}
\newcommand{\mstsn}{$M_{\star{\rm ,sn}}$}
\newcommand{\Mstsn}{M_{\star{\rm ,sn}}}
\newcommand{\dmrh}{$\dot{M}_{\rm rh}$}
\newcommand{\Dmrh}{\dot{M}_{\rm rh}}
\newcommand{\derh}{$\dot{E}_{\rm rh}$}
\newcommand{\Derh}{\dot{E}_{\rm rh}}
\newcommand{\dmbh}{$\dot{M}_\bullet$}
\newcommand{\Dmbh}{\dot{M}_\bullet}

\title[Feedback from quasars and galactic winds]
{Feedback from quasars in star-forming galaxies  \\
and the triggering of massive galactic winds}
\author[Monaco \& Fontanot]{Pierluigi Monaco \& Fabio Fontanot\\
Dipartimento di Astronomia, Universit\`a di Trieste, 
via Tiepolo 11, 34131 Trieste, Italy - email: monaco, fontanot@ts.astro.it}

\begin{document}

\date{Accepted ... Received ...}

\pagerange{\pageref{firstpage}--\pageref{lastpage}} \pubyear{2004}

\maketitle

\label{firstpage}

\begin{abstract}
The shining of quasars is a likely trigger of massive galatic winds,
able to remove most interstellar medium (ISM) from a star-forming
spheroid.  However, the mechanism responsible for the deposition of
energy into the ISM is still unclear.  Starting from a model for
feedback in galaxy formation with a two-phase medium (Monaco 2004a),
we propose that the perturbation induced by radiative heating from a
quasar on the ISM triggers a critical change of feedback regime.
In the feedback model, supernova remnants (SNRs) expanding in the hot
and pressurized phase of a star-forming spheroid tipically become
pressure-confined before the hot interior gas is able to cool.  In
presence of runaway radiative heating by a quasar, a mass flow from
the cold to the hot phase develops; whenever this evaporation flow is
significant with respect to the star-formation rate, due to the
increased density of the hot phase the SNRs reach the point where
their interior gas cools before being confined, forming a thick cold
shell.  We show that in this case the consequent drop in pressure
leads quickly to the percolation of all the shells and to the
formation of a super-shell of cold gas that sweeps the whole galaxy.
Radiation pressure is then very effective in removing such a shell
from the galaxy.  This self-limiting mechanism leads to a correlation
between black hole and bulge masses for more massive bulges than
$10^{10}$ \msun.

The insertion of a motivated wind trigger criterion in a hierarchical
galaxy formation model shows however that winds are not necessary to
obtain a good black hole--bulge correlation.  In absence of winds,
good results are obtained if the mechanism responsible for the
creation of a reservoir of low-angular momentum gas (able to accrete
onto the black hole) deposits mass at a rate proportional to the
star-formation rate.  Using a novel galaxy formation model, we show
under which conditions black hole masses are self-limited by the wind
mechanism described above, and outline the possible observational
consequences of this self-limitation.

\end{abstract}
\begin{keywords}
galaxies: active - galaxies: bulges - intergalactic medium - quasars: general
\end{keywords}

\section{Introduction}

Active Galactic Nuclei (AGN) are intimately connected to the
spheroidal components of galaxies.  This is highlighted at low
redshift by the correlation between the mass of the dormant black
holes hosted in ellipticals and spiral bulges and their mass or
central velocity dispersion (Kormendy \& Richstone 1995; Magorrian et
al. 1998; a more recent determination is given, e.g., by Marconi \&
Hunt 2003 or H\"aring \& Rix 2004).  The mass function of these black
holes is found to be consistent with that inferred from the accretion
history of quasars (Salucci et al. 1999; Yu \& Tremaine 2002; Shankar
et al. 2004; see also Haiman, Ciotti \& Ostriker 2004).  At higher
redshift, quasars and radio galaxies are systematically found to be
hosted at the centres of elliptical galaxies (see, e.g., Dunlop et
al. 2003).

These pieces of evidence point to a connection between the formation
of the two classes of objects.  Many authors have proposed that
feedback from the quasar could self-limit the bulge and black hole
masses, forcing them to be proportional (see, e.g., Ciotti \& Ostriker
1997; Silk \& Rees 1998; Haehnelt, Natarajan \& Rees 1998; Fabian
1999; Murray, Quataert \& Thompson 2005).  The dynamical importance of
feedback is confirmed by N-body hydro simulations (see, e.g.,
Springel, Di Matteo \& Hernquist 2004).  Besides, elliptical galaxies,
a homogeneous class of old, metal-rich, alpha-enhanced stellar
populations with little ISM, have longly been supposed to form through
a quick burst of star formation followed by a strong wind, able to
wipe the galaxy out of its ISM and to expel metals (especially iron)
to the inter-cluster medium (ICM) (see, e.g., Renzini 2004).  To
reproduce the correlation between stellar mass and the level of alpha
enhancement, this wind must halt star formation earlier in more
massive galaxies (Matteucci 1994, 1996).  Under this ``monolithic''
hypothesis, and using the shining of quasars as a clock for the
formation of elliptical galaxies, Granato et al. (2001; 2004) succeded
in reproducing the main statistical properties of elliptical galaxies
and quasars.  Also the high level of chemical enrichment in quasars
(see, e.g., Hammann \& Ferland 1999) is well explained in this context
(Matteucci \& Padovani 1993; Romano et al. 2003).

On the other hand, many galaxy formation models based on hierarchical
clustering are also successful in predicting the quasar--bulge
connection, even in absence of an explicit self-limiting mechanism, by
assuming simply that some fraction of cold bulge gas is accreted onto
the black hole (Kauffmann \& Haehnelt 2000; Cattaneo 2001; Cavaliere
\& Vittorini 2002; Hatziminaoglou et al. 2003; Mahmood, Devriendt \&
Silk 2004; Bromley, Somerville \& Fabian 2004). However, the level of
alpha-enhancement of stars in ellipticals is difficult to obtain in
this framework (Thomas 1999).

The energy budget of an accreting black hole radiating with an
efficiency $\eta=0.1$ amounts to $\eta M_\bullet c^2\simeq1.8\times
10^{61} (M_\bullet/10^8\ {\rm M}_\odot)$ erg.  A $M_\bullet \simeq
1.6\times10^8$ \msun\ black hole, which radiates $2.9\times10^{61}$
erg during his life, is typically hosted by a $M_{\rm bul}\simeq
10^{11}$ \msun\ bulge (see, e.g., Shankar et al. 2004). The binding
energy of such a bulge (with velocity dispersion $\sigma\simeq200$
\kms) is of order $M_{\rm bul}\, \sigma^2\simeq 8.0\times10^{58}
M_{\rm bul,11}\, \sigma_{200}^2$ erg (here $M_{\rm bul,11}=M_{\rm
bul}/10^{11}$ \msun\ and $\sigma_{200}=\sigma/200$ km s$^{-1}$).  It
suffices then to inject \circa0.3 per cent of the energy budget into
the ISM of a forming galaxy to influence it strongly, e.g. to trigger
a strong wind able to remove most ISM from the galaxy.

The mechanism responsible for this injection of energy is however
unclear.  In fact, a great part of the energy is emitted as UV-X
radiation.  Following Begelman (2004), a relatively hard UV-X source
can affect the ISM by two main processes.  The first is radiation
pressure, exerted especially on dust grains, that can push matter in
the radial direction.  However, its efficiency in accelerating matter
is of order $v/c$, where $v$ is the bulk velocity of the gas in the
radial direction.  Its effectiveness in creating bulk motions of gas
is then restricted to the neighbourhood of the central engine, where
the $r^{-2}$ dependence makes it very strong, thus compensating for
the very low initial $v/c$.  Recently, Murray et al. (2005) proposed
that radiation pressure alone can drive a rather robust wind, able to
remove some 10 per cent of the matter out of a bulge.  This is true if
the whole ISM is optically thick to radiation; we will come back to
this point in Section 3.

Radiation pressure is also a good candidate for causing the strong
outflows seen in Broad Absorption Line (BAL) quasars.  These objects,
relatively rare at the peak of quasar activity but rather common at
high redshift (see, e.g., Maiolino et al.  2003), show outflows along
the line of sight with velocity up to $\sim 0.1c$.  The kinetic energy
associated to these outflows is likely high; however, energy budget
and effectiveness in triggering a galaxy-wide wind depend sensitively
on the covering factor of the expelled gas.  BAL quasars could
correspond to a particular stage in which the quasar emits a very
strong wind with a high covering factor; this stage should last much
less than an Eddington time ($t_{\rm ed}\simeq 4\times10^7$ yr for
$\eta=0.1$).  Such an outflow would be effective in removing most ISM
from the host galaxy (see, e.g., Granato et al. 2004).  If instead BAL
quasars correspond to rather common, low covering factor events, their
ejection will most likely trigger star formation (similarly to what
happens in radio galaxies, where radio jets are aligned with
star-forming regios) more than quenching it by removing all the ISM.

Radiative heating is the second mechanism.  Quasars emit UV-X light
with a relatively hard spectrum, corresponding to inverse Compton
temperatures of order $T_{\rm IC}\simeq2\times 10^7$ K (see Sazonov,
Ostriker \& Sunyaev 2004b).  Assuming thermal equilibrium in presence
of this heating source, cold gas will be partially or totally heated
to a temperature $\sim T_{\rm IC}$ if radiation pressure is important
with respect to the thermal one of the ISM (Krolik, McKee \& Tarter
1981; Begelman, McKee \& Shields 1983).  Begelman (1985) estimated
that such ablation could evacuate the inner region of Seyfert
galaxies.  Moreover, radiation pressure can accelerate cold clouds to
some 100 km s$^{-1}$.  In a typical star-forming spheroid, whose ISM
is likely characterized by a high thermal pressure, radiative heating
will perturb the ISM in the inner region of the galaxy, but taken
alone it will not be able to cause a massive wind.  Sazonov et
al. (2004a) computed the effect of such heating on the ISM of a
forming spheroid; in their calculation a wind is triggered when the
fraction of old gas to total baryons is $\la1$ per cent, and star
formation is almost over.  As a result, the amount of ejected matter
is modest.

The radiation pressure of a black hole accreting at a rate \dmbh\ at a
distance $r$ is (assuming an efficiency $\eta=0.1$ as above and
expressing the pressure in terms of $P/k$, where $k$ is the Boltzmann
constant):

\begin{eqnarray} \label{eq:radpres} 
\frac{P_{\rm rad}}{k}&=&\frac{L}{4\pi r^2 ck} \\
&\simeq& 4.4\times10^7\ 
\left(\frac{\Dmbh}{4\ {\rm M}_\odot\ {\rm yr}^{-1}}\right) 
\left(\frac{r}{1\ {\rm kpc}}\right)^{-2} \ {\rm K}\ {\rm cm}^{-3} \, .
\nonumber\end{eqnarray}

\noindent
This relation is calibrated on the Eddington accretion rate (4
\msunyr) of a $1.6\times 10^8$ \msun\ black hole hosted in a $10^{11}$
\msun\ bulge.  For a star-forming spheroid the thermal pressure
$P_{\rm th}/k$ can be as high as $10^5-10^6$ K \cmt\ (see Section 2).
The limit for the existence of a hot phase subject to runaway heating
is $P_{\rm rad}/P_{\rm th}\ga 7-27$ (Begelman et al. 1983).  This is
valid for plain inverse-Compton heating; heating by metal line
absorption is very likely to contribute significantly (Ostriker \&
Ciotti 2004).  Taking a condition $P_{\rm rad}\ge 10 P_{\rm th}$,
runaway heating will be effective within a radius:

\be R_{\rm rh}=2.1
\left(\frac{\dot{M}_\bullet}{4\ {\rm M}_\odot\ {\rm yr}^{-1}} \right)^{1/2} 
\left(\frac{P_{\rm th}/k}{10^6\ {\rm K}\ {\rm cm}^{-3}} \right)^{-1/2}
\ {\rm kpc} \, .
\label{eq:r_rad} \ee

\noindent
This radius is of the same order of magnitude as the limiting radii
reported by Sazonov et al. (2004a).

An element that has not been considered in the literature is that the
perturbation induced by radiative heating will influence the way in
which feedback from SNe works.  In particular, radiation pressure can
be very efficient in driving a massive wind if some other mechanism is
able to generate a reasonably fast, optically thick outflow with a
high covering factor.  This can be provided by the percolation of many
SN remnants in the so-called Pressure-Driven Snowplough (PDS) phase,
that takes place after part of the internal hot gas of the remnant has
cooled and collapsed into a dense, cold shell (called ``snowplough'',
see, e.g., Ostriker \& McKee 1988).  In a previous paper (Monaco
2004a, hereafter paper I), we presented a model for feedback in galaxy
formation.  This model, which is described in Section 2, predicts that
the ISM of a typical star-forming spheroid is characterized by a very
hot and pressurized gas phase; this is able to halt the expansion of
SN remnants much earlier than the time required by the internal gas to
cool; the resulting SNRs are characterized by a very low porosity.  We
show in this paper that the insertion of a physically motivated
evaporation term induced by the quasar leads very naturally (and
without any tuning of the parameters) to a critical change of the
feedback regime that causes, through the percolation of snowploughs,
the creation of an optically thick super-shell that expands out of the
galaxy at a speed of \circa200 km s$^{-1}$.  Such a shell, pushed by
radiation pressure, can be ejected out of a large spheroid, quenching
star formation and leaving behind a hot rarified bubble.  This
mechanism can lead to a self-regulated black hole--bulge relation very
similar to the observed one.

As mentioned above, the black hole--bulge relation can be generated
simply by the mechanism responsible for the nearly complete loss of
angular momentum, necessary to the gas to be able to accrete onto the
black hole.  A good fit is obtained if the accumulation of a reservoir
of low-angular momentum gas proceeds at a rate proportional to the
star formation rate in the bulge, or equivalently if a fraction of the
cold bulge gas is systematically put into this reservoir.  To assess
the actual importance of winds in hierarchical galaxy formation, we
insert a motivated wind criterion into a new galaxy formation model
(that follow losely the steps of commonly used ``semi-analytic''
models), and show with an example under which conditions the black
hole mass is self-limited by winds more that determined by the
reservoir mechanism.

The paper is organized as follows.  Section 2 gives a very short
summary of the model for feedback presented in paper I.  Section 3
shows the effect of the introduction of a physically motivated
evaporation term and discusses the triggering of the wind.  In section
4 quasar-triggered winds are introduced in the galaxy formation model.
Section 5 the conclusions.  Preliminary results were shown by Monaco
(2004c).

\section{The model for feedback}

We give here a brief description of the model for feedback presented
in paper I; please refer to that paper for all details, and for a
complete discussion of all the hypotheses.

Consider a volume $V$ filled with a two-phase ISM, made up by a
pervasive hot phase, of density and temperature \nh\ and \th, and a
distribution of cold clouds, of density and temperature \nc\ and \tc.
Pressure equilibrium is assumed, so that $\Nh\Th=\Nc\Tc$.  Fixing the
cold phase temperature to \tc=100 K, it is very easy to express the
filling factors of the two phases in terms of their temperatures and
mass fraction (equation 2 of paper I).

Cold clouds are assumed to have a power-law mass function,
$n(m\cl)dm\cl\propto m\cl^{-\alpha\cl}dm\cl$.  The parameter
$\alpha\cl$ is set here for simplicity to $-2$, the expected value for
a typical fractal distribution (Elmegreen 2002).  The range of allowed
masses is bounded below by a mass $m_{\rm l}=0.1$ \msun, and above by
a mass $m_{\rm u}$.  The upper bound is fixed as follows: all clouds
more massive than the Jeans mass $m_{\rm J}$ (that takes into account
the non-sphericity of clouds, see Lombardi \& Bertin 2001) collapse in
a dynamical time \tff, form stars and are finally destroyed within
another \tff.  During the $\sim 2t_{\rm dyn}$ period they acquire mass
through unelastic collisions with smaller clouds, up to a mass $m_{\rm
u}$, computed with the aid of the Smoluchowski equation of kinetic
aggregations.  The fraction \fcoll\ of cold gas involved in the star
formation process is then equal to the fraction of mass in clouds with
$m_{\rm J}\le m\cl\le m_{\rm u}$.

A fraction \fstar=0.1 of the collapsing cloud is transformed into
stars.  In Monaco (2004b) we showed that, due to the multi-phase
nature of the ISM, the explosion of SNe leads to the destruction of
the star-forming cloud in \circa3 Myr, leading to a modest loss of
\circa5 per cent of the energy budget.  SN remnants percolate very
soon, creating a single SB for each forming cloud.  The SBs expand
according to the solution of Weaver et al (1977); they propagate into
the hot pervasive phase, their (modest) interaction with the cold
clouds is neglected.  The fate of SBs depends on the vertical
scale-length of the system they belong to: if the system is thin
enough, like in spiral discs, SBs blow out of the system before they
are pressure-confined by the hot phase.  In this case the efficiency
of feedback in heating the local ISM (i.e. the fraction of energy
given to the hot phase) is $f_E\sim5-10$ per cent, while the remaining
energy is injected into the external halo.  For thicker systems, like
a star-forming spheroid, SBs end by being confined by the pressure of
the hot phase.  Both blow-out and pressure confinement can take place
either in the adiabatic expansion phase, where the thermal energy of
the hot internal gas of the SB is conserved, or after that part of the
internal gas has cooled down and collapsed to a cold shell, pushed by
the pressure of the remaining hot gas (the PDS).  When the PDS forms,
the hot phase is not shock-heated any more but collapsed into the
snowplough, so its net effect for the ISM is a loss of thermal energy.

Four different self-regulated feedback regimes can arise, depending
on whether the SBs end by blow-out or pressure confinement, before or
after the PDS stage starts.  While blow-out in the PDS stage is not
common, blow-out in the adiabatic stage is the typical feedback regime
in galaxy discs.  In a Milky Way-like case, the main properties of our
local ISM are recovered.  Adiabatic confinement is the typical regime
found in thick system, while in very dense thick systems SBs are
confined in the PDS regime.  In some critical cases the hot phase is
strongly depleted and the cold phase percolates the volume, giving
thus rise to a burst of star formation.

In the following we concentrate on thick systems.  In the adiabatic
confinement regime, the details of the SB evolution are not important,
as all the energy that manages to escape the star-forming cloud is
given to the hot phase.  

The system considered in paper I is composed by cold and hot gas
phases, stars and an external halo (a passive reservoir of gas).  Cold
gas flows into the volume from the external halo on some specified
infall time \tinf:

\be \Dminf = \frac{\Mhalo}{\Tinf}\, . \label{eq:dminf} \ee

\noindent
Star formation proceeds at a rate:

\be \Dmsfr = \Fstar \Fcoll \frac{\Mcold}{\Tff}\, , \label{eq:dmsfr} \ee

\noindent
and enriched gas is restored from dying stars (in an instantaneous
recycling approximation) at a rate:

\be \Dmrest = \Frest \Dmsfr\, , \label{eq:dmrest} \ee

\noindent
where \frest\ is the fraction of restored mass (assumed here to be 0.2).
Besides, a fraction \fevap=0.1 of the star-forming cloud is evaporated
by HII regions and SNe (see Monaco 2004b), giving origin to an
evaporation flow:

\be \Dmev = \Fevap \Fcoll \frac{\Mcold}{\Tff}\, . \label{eq:dmev} \ee

\noindent
The hot phase cools at a rate:

\be \Dmcool = \Fcool \frac{\Mhot}{\Tcool}\, , \label{eq:dmcool} \ee

\noindent
where the cooling time \tcool\ is computed using the simple
approximation proposed by Cioffi, McKee \& Bertschinger (1988) and
\fcool=0.1 is a parameter that regulates the passage from the hot to
the cold phase (whose value depends on the complex density structure
of the cooling gas).  Finally, the hot phase typically is not confined
within the star-forming region (the volume $V$ that contains the ISM),
so it leaks out at a rate:

\be \Dmleak = \frac{\Mhot}{\Tleak}\, , \label{eq:dmleak} \ee

\noindent
where the leaking time \tleak\ is the sound-crossing time of the
structure.

The mass of the four components evolves according to this set of
equations:

\be \left\{
\begin{array}{lcl}      
\Dmcold &=& \Dminf   +\Dmcool -\Dmsfr-\Dmev\\
\Dmhot  &=& -\Dmcool -\Dmleak +\Dmev +\Dmrest\\
\Dmstar &=& \Dmsfr - \Dmrest \\
\Dmhalo &=& -\Dminf + \Dmleak
\end{array} \right. \label{eq:massflows} \ee

\noindent
An analogous system can be easily written for the metal flows.

The hot phase gains thermal energy from SNe (both through blasts
and through the evaporated gas) at a rate:

\be \Desn = E_{51} \frac{\Dmsfr}{\Mstsn} \, , \label{eq:desn} \ee

\noindent
where \esn\ is the energy of the single SN in units of $10^{51}$ erg
(for simplicity we subtract from it any eventual loss due to the
destruction of the star-forming cloud) and one SN progenitor is formed
each \mstsn\ of stars (we take \mstsn=120 \msun).  Cooling and
leak-out lead to energy losses at rates:

\be \Decool = \frac{\Eh}{\Tcool}\, , \label{eq:decool} \ee

\be \Deleak = \frac{\Eh}{\Tleak}\, . \label{eq:deleak} \ee

\noindent
The evolution of the energy of the hot phase is then:

\be \Dehot = -\Decool -\Deleak +\Desn\, . \label{eq:nrgflux} \ee

In case of PDS confinement, the \desn\ term must take into account the
energy radiated away by the SB, while snowploughs generate a further
mass flow \dmsnpl\ from the hot to the cold phase, with the
corresponding loss of thermal energy \desnpl\ (see paper I for
details).

\begin{figure}
\centerline{
\includegraphics[width=8.4cm]{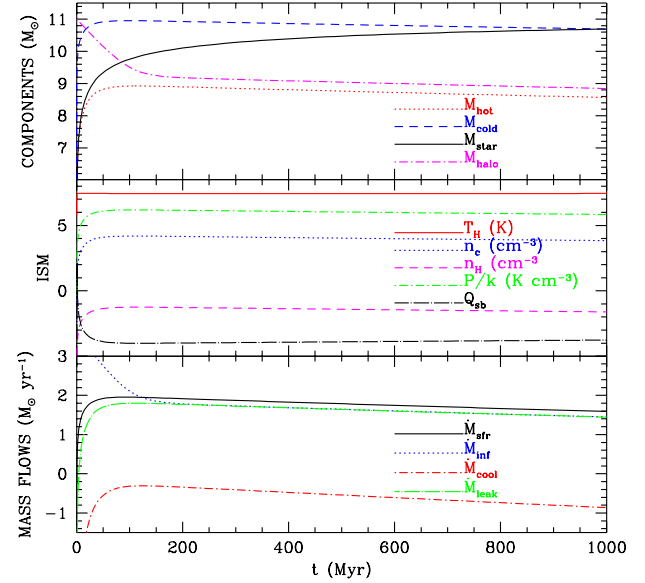}
}
\caption{Evolution of a toy spheroid with the feedback model of paper
I.  $10^{11}$ \msun\ are first put into the external halo and then let
infall on the ``galaxy''.  The upper panel shows the evolution of the
mass in the four components (see the labels) as a function of time.
The mid panel shows the evolution of the ISM (pressure, density of the
two phases, temperature of the hot phase, porosity of SBs).  The lower
panel shows the main mass flows. All quantities in the y-axes are
logarithmic.}
\label{fig:noqso}
\end{figure}

As an illustrative example we show the evolution of a ``monolithic''
spheroid of mass $10^{11}$ \msun\ and half-mass radius (used as
vertical scale-length) $R_{\rm hm}=4.9 kpc$.  Its circular velocity at
$R_{rm hm}$ is 209 \kms, and its average density at the same radius is
0.1 \msun\ pc$^{-3}$.  The infall time is assumed to be equal the
dynamical time, \tinf=$2.5\times10^7$ yr.  Figure 1 shows the
evolution of the four mass components, the main mass flows and the
state of the ISM as a function of time.  The cold gas accumulates
quickly, but the efficient feedback prevents stars from forming as
quickly.  The ISM self-regulates to $P/k$\circa$10^6$ K \cmt,
\th\circa$2.5\times10^7$ K, \nh\circa$4\times10^{-2}$ \cmt\ and
\nc\circa$10^4$.  Star formation regulates to a value of \circa50
\msunyr, while infall and leak-out assume very similar values.
Cooling is always negligible.  Most importantly, the porosity \poro\
of the expanding SBs is always very low, as the blasts are very
quickly halted by the high pressure of the external gas.

The feedback regime described above applies to the case when the cold
gas infalls smoothly to the galaxy in small chunks, so that the
creation of collapsing clouds is left to the coagulation mechanism
introduced above.  This is not the case in major mergers of disc
galaxies; tidal disturbances in the last phases of the merger increase
the thickness of the system, that switches then from adiabatic
blow-out to adiabatic confinement.  The sudden decrease of the Jeans
mass, due to the increased pressure, adds then to the tidal
disturbances in causing a diffuse burst of star formation on a
dynamical timescale.  However, only a fraction \fstar\ of each cloud
is transformed into stars, so this burst will contribute to the
increase of pressure but will not consume a very large amount of
stars.  Tidal disturbances, besides helping the triggering of star
formation in clouds, will also stretch and fragment them in small
pieces, helping the system to get into the new regime.  Most gas will
then be transformed into stars when the new feedback regime has taken
place.

A more thorough description of this process and a more careful
comparison to observations is left to a forthcoming paper.  Here we
notice only that the adiabatic confinement regime implies the presence
of significant amounts of molecular gas (with such a high \nc\
molecular hydrogen should be present even in non-collapsing clouds) in
conjunction with a relatively modest star formation activity.  This is
nicely consistent with the observation of cold gas in elliptical
galaxies at the centre of massive cooling-flow clusters (see, e.g.,
Edge \& Frayer 2003).

\begin{figure*}
\centerline{
\includegraphics[width=8.4cm]{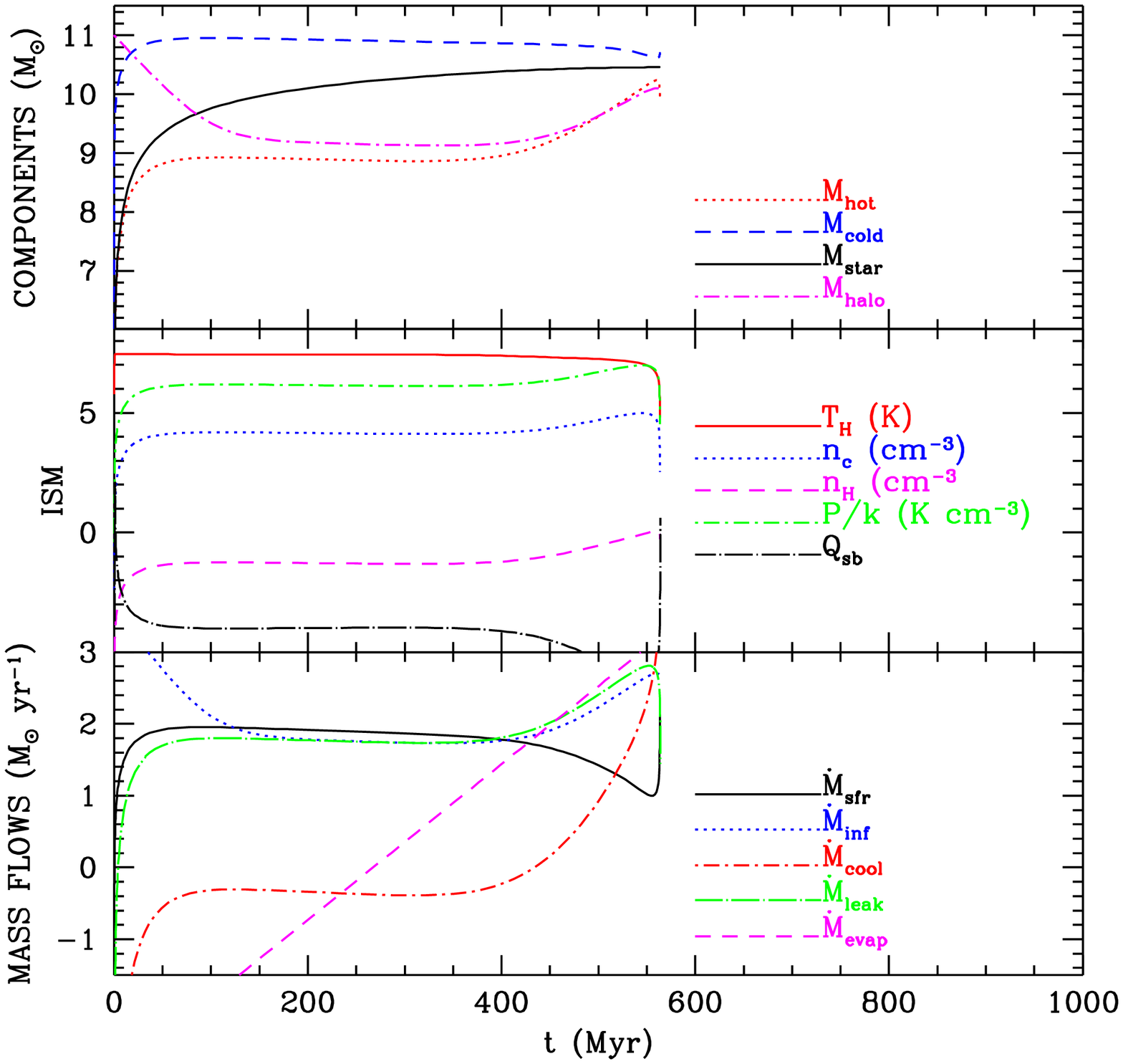}
\includegraphics[width=8.4cm]{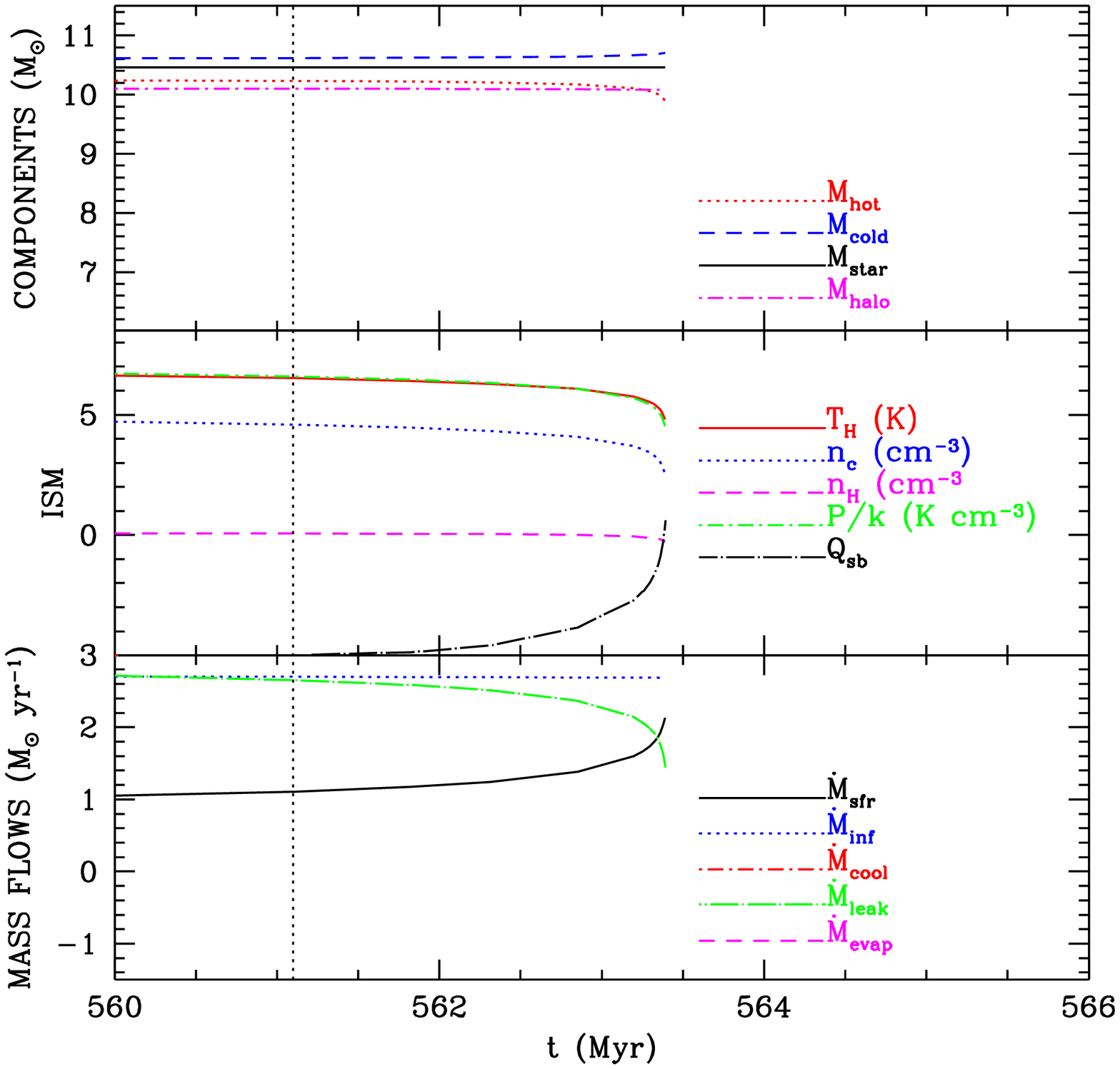}
}
\caption{As in Figure~\ref{fig:noqso} for the evolution of the toy
spheroid in case of runaway radiative heating.  The right panel shows a detail
of the critical change of regime; the dotted vertical line denotes the
starting time of the PDS confinement regime.}
\label{fig:qso}
\end{figure*}

\section{Feedback from the quasar}

\subsection{The effect of radiative heating}

We assume for the moment that the whole galaxy is affected by
radiative heating in the same way. To model the perturbation induced
by this heating we add to the system described above a seed black hole
of 1000 \msun, accreting mass at the Eddington rate for the whole
period.  We model radiative heating as an evaporation term \dmrh,
proportional to the accretion rate onto the black hole \dmbh, that
moves mass from the cold to the hot phase:

\be \Dmrh = \Krh \Dmbh \, . \label{eq:rh} \ee

\noindent
This term gives a positive contribution to the \dmhot\ equation and a
negative one to the \dmcold\ equation in the
system~\ref{eq:massflows}.  The corresponding energy flow term, \derh,
is obtained assuming that the gas is heated to the inverse Compton
temperature of the AGN, $T_{\rm IC}=2\times 10^7$ K:

\be \Derh = \frac{3}{2}kT_{\rm IC} \frac{\Dmrh}{\mu m_p} 
\label{eq:derh} \ee

\noindent
(here $\mu$ is the mean molecular weigth of the hot phase, and is
self-consistently computed in the model). The global heating rate of
the quasar is $6.7\times 10^{44}\tau T_{\rm IC8}L_{46}$ erg s$^{-1}$
(Begelman 1985), where $\tau$ is the electron scattering optical depth
of the cold phase, $T_{\rm IC8}=T_{\rm IC}/10^8$ K and $L_{46}$ is the
ionizing radiation of a quasar in units of $10^{46}$ erg.  The optical
depth can be written as that of a single cloud, $\tau_{\rm cl}$, times
the covering factor ${\cal C}$ of the cold phase.  For a typical AGN
spectrum a fraction $\Fion\simeq37$ per cent of the radiated energy is
above the Lyman limit\footnote{This is computed using the quasar
template spectrum of Cristiani \& Vio (1990) down to 538\AA,
extrapolated to 300 \AA\ with the recipe of Risaliti \& Elvis (2004).
At shorter wavelenght (between 0.01 and 30 \AA) we use a power-law
with a photon index of $\Gamma = -1.8$ (see Comastri et al., 1995).
The relative normalization is fixed by assuming $\alpha_{ox} = -1.63$
(Vignali et al. 2003).  The
interpolation between 30 and 300 \AA\ follows Kriss et al. (1999).}.
In this case $L_{\rm ion}=2.0\times10^{45} \Dmbh$ erg s$^{-1}$, where
\dmbh\ is in units of \msunyr.  The constant \krh\ of
equation~\ref{eq:rh} is then estimated by equating the heating rate to
the thermal energy gained by the evaporated mass.  We obtain:

\be \Krh = 104 \tau_{\rm cl} {\cal C}\, . \label{eq:krh} \ee

\noindent
This is likely to be a lower limit, due to the neglected metal line
heating.  The cold phase of a highly pressurized ISM like that of
Figure~1, with the mass subdivided into many small clouds, presents a
rather high covering factor\footnote{We have ${\cal C}=3f_cR/4a_{\rm
cl}$, where $f_c$ is the filling factor of the cold phase, $R$ the
size of the structure and $a_{\rm cl}$ the typical size of the cold
clouds.  For the model in Figure~1, $f_c\sim2\times10^{-4}$ and
$a_{\rm cl}\sim1$ pc, so that for $R\sim2$ kpc (the size at which a
typical $10^{11}$ \msun\ bulge with an Eddington-accreting black hole
is heated) ${\cal C}\sim0.3$.}, ${\cal C}$\circa 0.3.  The value of
$\tau_{\rm cl}$ could be low for these very high density clouds.
However, in a more realistic setting a significant fraction of the
cold phase would be in a warm intermediate phase, with $T\sim10^4$.
While a modeling of the dynamical role of this warm phase is beyond
the interest of this paper, it is clear that this phase would present
significant values of $\tau_{\rm cl}$ and ${\cal C}$.  Heating by metal
lines would obviously strenghten this conclusion.  On the other hand,
attenuation by diffuse dust would give lower evaporation rates.  As a
conclusion, we reckon that a value of \circa20-50 for \krh\ is
reasonable.  Begelman (1985, 2004) quotes an evaporation rate of
$20-200\,{\cal C}(L_{\rm ion}/10^{46}\ {\rm erg})$ \msunyr\ for a
spiral-like ISM.  This would correspond to $\Krh=4-40\,{\cal C}$, lower
than the value quoted above.  But Begelman's calculation refers to the
rather different case of a wind generated in a spiral galaxy by a
central AGN.  In our case the geometry of the system is rather
different, and we do not require that the gas is ejected out of the
galaxy in a wind.  This justifies the higher evaporation rate.

Figure~2 shows the effect of this heating term on the system.  As soon
as the radiative heating mass flow becomes comparable to the other
ones (\dmsfr, \dmleak\ and \dminf), \circa400 Myr after the beginning,
the density of the hot phase starts increasing.  The pressure grows,
while the temperature of the hot phase starts to decrease, due to the
enhanced cooling.  SBs are confined by the enhanced pressure earlier
and earlier, and \poro\ decreases considerably.  When $n_h$ grows
large enough, the system switches to the PDS confinement feedback
regime.  In this case a significant amount of energy from SNe is
radiated away before the blast is pressure-confined by the hot phase.
The resulting snowploughs move mass from the hot to the cold phase,
causing a sudden decrease of the density and temperature of the hot
phase.  This leads to a drop in thermal pressure, with a consequent
increase of the Jeans mass of cold clouds and of the star formation
rate.

Most importantly, just after the change to the PDS confinement regime
has taken place, the porosity of the SBs jumps suddenly from very low
to $>1$ values.  This means that the SBs percolate into a unique
super-SB, a cold super-shell that sweeps the galaxy, cleaning it very
efficiently from all of its ISM.  The final velocity of SBs at the
formation time of the super-shell will not be much higher than the
thermal velocity of the hot phase, \circa200 \kms.  This means that
the outgoing mass will be accelerated by radiation pressure with an
initial efficiency of \circa0.06 per cent.  This efficiency will grow
as the shell is accelerated to a higher speed.

The percolation of a diffuse distribution of blasts will produce not
only an outgoing supersonic super-shell.  A significant fraction of
the mass will be compressed to the centre of the galaxy by the same
blasts.  This is well illustrated, for instance, by the simulation of
Mori, Ferrara \& Madau (2002) of a blow-out in a primordial galaxy.
Though the physical scales and the energetic budget involved are quite
different, the geometry and the astrophysics of the problem are
similar.  In their simulations a fraction of order of 10 per cent of
the gas (the precise number depending on the distribution of the
star-forming clouds) is compressed to the centre of the halo, while
the rest is blown away.  In our case, the gas compressed to the centre
will give rise to a secondary burst of star formation and to further
accretion onto the black hole.  This will eventually give more energy
to the super-shell, both in terms of new SNe giving thermal energy to
the rarified hot gas and in terms of new radiation pressure from the
accreting black hole.  Moreover, if the ISM is dusty in the starburst
phase, the dust will be destroyed by the quasar or removed by the wind
just after the percolation phase.  The main unobscured shining phase
of the quasar could then correspond to the accretion of the matter
compressed to the centre by the blasts.

The super-shell will form mainly by the sweeping of the hot phase,
which at the onset of PDS confinement amounts to \circa20 per cent of
the mass, while the very dense cold clouds (\circa50 per cent of the
mass) will be affected less strongly.  So, the super-shell may leave
some cold mass behind.  However, it is likely that this mass will be
promptly consumed into stars, for the following reasons.  Radiative
heating acts preferentially on cold clouds, so the matter left will be
in relatively large clouds, where star formation is easier to trigger.
At the super-shell formation, sweeping by the supersonic shells will
trigger star formation in many of the remaining clouds.  After
the formation of the super-shell, the galaxy will be filled with hot,
pressurized gas, able to thermo-evaporate the leftovers of the star
formation process (which amount to \circa90 per cent of the mass
involved in star formation).  Moreover, radiative heating and
radiation pressure themselves can pressurize the clouds, stimulating
star formation (Begelman 1985).  As a conclusion, it is likely that
the cold gas left behind by the shell will be promply consumed into
stars or evaporated.

From this toy model we learn that the galactic wind is triggered
whenever the evaporation rate is roughly a factor of 10 larger than
the (unperturbed) star formation rate.  As this quantity scales with
the initial gas mass, and as the Eddington accretion rate is
proportional to the black hole mass, this ``monolithic'' toy model
implies a correlation between stellar and black hole masses, due to
the self-limiting feedback from the quasar.  However, the black hole
mass at the shining time is \circa$2\times 10^9$ \msun, much larger
than what expected in this spheroid.  Moreover, no mention is done
here on how gas manages to lose angular momentum so as to be able to
accrete onto the black hole.

\subsection{Radiation pressure on the expanding super-shell}

The mass that can be accumulated initially in the super-shell is
limited by the kinetic energy that the shell can receive from SNe.
Figure~2 shows that at the percolation time the star formation is as
high as \circa100 \msunyr.  For \mstsn=120 \msun, this corresponds to
0.83 SN per year.  This level of star formation will be sustained for
at least one shell-crossing time, which is the time required to the
shell to form.  For a fiducial initial velocity $v_i=200$ \kms,
coincidentally very similar to the circular velocity of a $10^{11}$
\msun\ bulge at the half-mass radius ($R_{\rm hm}=4.9$ kpc), $V_c=209$
\kms, the shell crossing time is very similar to the dynamical time of
the bulge, $2.5\times 10^7$ yr, and also to the time required to an
8-\msun\ star to explode.  In this time about $2.2\times 10^7$ SN
progenitors are formed, for an energy budget of \circa$2\times
10^{58}$ erg.  The kinetic energy of a $10^{10}$ \msun\ shell
(amounting to 10 per cent of the total mass) traveling at 200 \kms\ is
\circa$4\times10^{57}$; it will be accelerated by SNe if a reasonable
20 per cent efficiency is assumed.  However, this is likely to be only
a lower limit.  Indeed, such a starburst consumes only $2.5\times10^9$
\msun\ of stars.  In other words, \circa4 \msun\ are accelerated for
each \msun\ of stars formed.  As commented above, the formation of the
supershell will likely trigger a burst of star formation, both by
compressing cold gas to the centre and by sweeping the most massive
cold clouds.  If for instance $10^{10}$ \msun\ of gas are transformed
into stars during the formation of the super-shell, the fraction of
mass that can be accelerated by SNe raises to 40 per cent.  We
conclude that the available energy from SNe does not put strong
constraints on the amount of mass that can be accelerated.

The amount of mass that can be ejected in a wind is instead limited by
the ability of radiation pressure to perform work on the shell.  Due
to its high pressure, the gas in the shell is not subject to runaway
radiative heating until it becomes subsonic or is destroyed by
Raileigh-Taylor instabilities.  While sweeping the hot gas phase and,
later, the hot halo gas pervading the dark matter halo that surrounds
the galaxy, the super-shell grows in mass.  It is slowed down by
gravity, pressure from the external hot gas and mass load.

The equation of motion of a super-shell of mass $M_s$, radius $R_s$
and velocity $v_s$ can be written as follows:

\begin{eqnarray}\label{eq:shell} 
\lefteqn{\frac{1}{2}M_s\frac{dv_s^2}{dR_s} =} \\
&&-v_s^2\frac{dM_s}{dR_s}
- GM_s\frac{M_{\rm tot}(R_s)}{R_s^2} + (P_{\rm rad}+P_{\rm int}-P_{\rm ext})
4\pi R_s^2 \, . 
\nonumber\end{eqnarray}

\noindent
We will neglect the internal pressure term, $P_{\rm int}$, and
concentrate on radiation pressure (equation~\ref{eq:radpres}).

Let's first assume that a shell of constant mass is leaving an
isolated bulge of mass $M_{\rm bul}$; no dark matter component is
considered at the moment.  Fitting the data of Marconi \& Hunt (2003),
we find that the half-mass radius ($R_{\rm hm}=1.35R_e$, where $R_e$
is the effective radius) of an elliptical galaxy scales with the bulge
mass as follows:

\be R_{\rm hm}=4.9 M_{\rm bul,11}^{0.65}\ {\rm kpc} \, . \label{eq:rhm} \ee

\noindent
The scatter around this relation is \circa0.3 dex.  For the mass
profile we assume for simplicity a roughly constant rotation curve, so
that $M_{\rm bul}(r)\propto r$.  We then fix the value of $M_{\rm
bul}(r)/r$ to its value at the half-mass radius, $M_{\rm bul}/2R_{\rm
hm}$.

The mechanical pressure exherted by radiation depends on the optical
depth of the shell, $P_{\rm rad}(1-{\rm e}^{-\tau_{\rm shell}})$.  The
column density of a shell of mass $M_s=f_s M_{\rm bul}$ at a radius
$R$ (in kpc) is $8.0\times10^{23}\, f_s\, M_{\rm bul,11} R^{-2}\ {\rm
cm}^{-2}$.  Following Begelman (2004), electron scattering gives only
$\tau_{\rm shell}\sim0.7 x\, f_s M_{\rm bul,11} R^{-2}$, where $x$ is
the ionization fraction.  Photoionization gives $\tau_{\rm shell} \sim
100\, (P_{\rm th}/P_{\rm rad})\, f_s M_{\rm bul,11} R^{-2}$.  As long
as the shell propagates supersonically, its pressure is very high, so
the value of $\tau_{\rm shell}$ at $P_{\rm th}=P_{\rm rad}$ is a very
conservative lower bound.  The high pressure guarantees also that the
shell is not affected by runaway radiative heating.  Dust absorption
on the other hand gives $\tau_{\rm shell}\sim 10^5 \, f_s M_{\rm
bul,11} R^{-2}$.  Moreover, the ratio between the mean free path of
the dust grains and the thickness of the shell results $\sim
4\times10^{-5} f_s^{-1} M_{\rm bul,11}^{-1} R^{2}$ (Murray et
al. 2005), so that dust grains are hydrodynamically coupled to the
fluid. As a conclusion, the shell absorbs most of the quasar light
from \circa4000 to \circa1 \AA, so that radiation pressure
(equation~\ref{eq:radpres}) can be safely used in
equation~\ref{eq:shell}.  This is valid until the shell is so diluted
that $\tau_{\rm shell}\sim1$, which happens only at several half-mass
radii.

Solving equation~\ref{eq:shell}, we find that the kinetic energy of
the shell, $K_s$, evolves like:

\be
K_s = -f_s M_{\rm bul}V_c^2 \ln\left(\frac{R_s}{R_i}\right)
+\frac{L}{c}(R_s-R_i) + K_i \, .
\label{eq:voidshell} \ee

\noindent
Here $V_c$ is the circular velocity of the bulge at $R_{\rm hm}$, and
the suffix $i$ refers to the initial conditions, so that
$K_i=f_sM_{\rm bul}v_i^2/2$.  Requiring that the minimum of $K_s$ is
positive, we obtain an upper limit on $f_s$ which is given by the
largest root of the algebraic equation $(a+\ln(b/f_s))f_s-b=0$, where
$a=1+v_i^2R_{\rm hm}/GM_{\rm bul}$ and $b=2R_{\rm hm}LR_i/GM_{\rm
bul}^2c$.  Scaling the accretion rate to 4 \msunyr\ (the Eddington
accretion rate of a $1.6\times 10^8$ \msun\ black hole hosted in a
$10^{11}$ \msun\ bulge) and assuming that the initial radius of the
shell is that at which radiation pressure equals the thermal one
(equation~\ref{eq:r_rad}), we obtain that the upper limit to $f_s$ is
fit (within \circa15 per cent) by the following formula:

\be
f_s<0.21
\left(\frac{\dot{M}_{\bullet,4}}
          {M_{\rm bul,11}^{1.1}}\right)^{1.5}\, .
\label{eq:f_s} \ee

\noindent
(here $\dot{M}_{\bullet,4}=\dot{M}_{\bullet}/4$ \msunyr).

Radiation pressure is then able to expel up to 21 per cent of the mass
of a star-forming spheroid if its black hole follows the known black
hole--bulge relation and is accreting at the Eddington ratio.  More
active black holes can remove much larger amounts of mass (assuming
that star formation is strong enough to create such massive
super-shells).  This provides a self-limiting mechanism able to
generate a black hole--bulge correlation very similar to that
observed\footnote{Equation~\ref{eq:r_rad} is used assuming a constant
gas pressure for all bulges.  Smaller bulges are however denser. From
the model described in Section 2 we find that roughly $P_{\rm
th}\propto M_{\rm bul}^{-0.5}$.  Taking this into account, we obtain
$f_s\la 0.2 (\dot{M}_{\bullet,4}/M_{\rm bul,11}^{0.9})^{1.5}$.}.

This result is valid for bulges with escape velocities larger that 200
\kms.  Less massive bulges than $10^{10}$ \msun\ have escape
velocities below this limit, so mass removal is efficient even in case
of no accretion.  This implies a much looser black hole--bulge
correlation at such masses.

Once the super-shell has left the galaxy, it interacts with the hot
gas pervading the dark matter halo.  In this case the gravity term of
equation~\ref{eq:shell} contains also the contribution of the halo,
while the hot halo gas slows down the shell by mass load and thermal
pressure.  We have verified that for reasonable values of the halo
parameters (mass, concentration, gas distribution etc.) the shell will
be promptly stopped as soon as it interacts with the gas.  At this
point it will fragment.  At a speed of \circa200 \kms, a distance of a
few tens of kpc is reached in several tens of Myr, i.e. \circa2-3
Eddington times.  If the accretion of the material compressed at the
centre started promptly at shell formation, and if the quasar were
visible only after the shell is destroyed, then most accretion would
be hidden by the dusty shell.  However, the time necessary to the gas
compressed to the centre to lose its angular momentum is likely to be
not negligible.  For instance, Granato et al. (2004) suggest a
timescale for viscous accretion of order $5\times10^7$ yr for a black
hole of $10^8$ \msun\ in a 200 km s$^{-1}$ bulge, amounting to
\circa1.2 Eddigton times.  We suggest as a likely scenario that while
the shell is pushed by accretion of the low-angular momentum gas
accumulated during the stage of self-limited star formation (in the
adiabatic confinement feedback regime), the accretion of the material
compressed at the centre at the super-shell formation takes place
mostly after the shell has been destroyed.

The drop in pressure consequent to the shell destruction makes
radiative heating effective again and, due to the very high covering
factor of the shell gas, a part of it can be in principle heated back
to $T_{\rm IC}$.  However, it is easy to verify that the heating time
evolves like $t_{\rm heat} \sim 1.2\times 10^8\ R^2 L_{46}^{-1}
(T/T_{\rm IC}) N_{H,24}$ yr, where $T$ is the temperature of the
heated gas and $N_{H,24}$ the column density of the layer that is
affected by heating, in units of $10^{24}$ cm$^{-2}$ (see Begelman
2004).  Clearly, for distances significantly larger that one kpc, the
heating time becomes too large for runaway heating to be effective. In
this case the shell will be heated to a much lower temperature than
$T_{\rm IC}$.

During the expansion, radiation pressure performs a work $L R_s/c$ on
the super-shell.  For $\dot{M}_\bullet=4$ \msun yr and $L=2.3\times
10^{46}$, at a fiducial distance of 20 kpc (roughly 5 half mass radii,
at which the effects of the dark-matter and hot-gas halo should be
important) the AGN has given $4.7\times10^{58}$ erg to the shell.  The
efficiency of energy injection depends on the actual espansion
velocity, which however does not exceed much the initial one if the
shell is near its mass limit.  Besides, for $k_{\rm rh}=50$ and
$M_\bullet=1.6\times10^8$ \msun, radiative heating injects
$6.5\times10^{59}$ erg into the ISM, amounting to 2 per cent of the
energy budget.  However, most of this energy is lost to radiation
during the formation of the super-shell.  This calculation confirms
that the energetic efficiency of feedback from quasars is very likely
smaller then 1 per cent.  In this case, the 5 per cent efficiency used
in the simulations of Springel et al. (2004) seems a rather extreme
choice.

To understand the robustness of these conclusions, we have repeated
the computation of the fraction of mass removed by the wind under
several different assumptions on the mass profile, obtaining in most
cases a fraction of order of 10 per cent times
$(\dot{M}_{\bullet,4}/M_{\rm bul,11})$ to the power 1.5--2 ($M_{\rm
bul,11}$ may be raised to a power very similar to one).

As a conclusion, radiation pressure can remove about 20 per cent of
the mass of a star-forming spheroid hosting an Eddington accreting
black hole that follows the local black hole--bulge relation.
However, it cannot remove this gas from the massive dark-matter halo
that hosts the bulge.

\subsection{Feedback from AGN in presence of other stellar feedback regimes}

The mechanism described above depends on the assumption that feedback
is in the adiabatic confinement regime when the quasar shines.  Here
we show that the validity of the numbers given above is more general.

As shown in Monaco (2004b), small star-forming clouds are destroyed by
a single SN.  As the number of SNe per cloud is rather small, the
fraction of energy lost in destroying the cloud could be rather high.
This justifies a lower effective value for \esn\ in the case of
adiabatic confinement.  In paper I it was found that, for \esn=0.3,
the system goes to the PDS confinement regime.  This has the effect of
lowering the pressure and increasing the porosity.  The onset of the
PDS confinement regime can be smooth (as the example given in paper I)
or critical; for the infall times used here the onset is critical,
leading to the percolation of SBs.  This will lead to critical,
self-stimulated bursts of star formation, that will increase the
pressure and make the solution bounce back to the adiabatic
confinement one.  In principle, this would lead to the creation of a
galaxy-wide super-shell.  However, in absence of a synchronized
trigger as a shining quasar, this explosion will presumably interest
different parts of the galaxy at different times; moreover, even in
case a super-shell is formed, its velocity will be lower than the
escape velocity of the cloud, so the cold gas will fall back soon if
$M_{\rm bul}>10^{10}$ \msun.  As a conclusion, the system will spend
most of its time in the adiabatic confinement regime, with possible
(self-stimulated and quickly self-quenched) bursts of star formation;
consequently, the effect of radiative heating will be similar to what
described above.

Murray et al. (2005) estimated that radiation pressure by an Eddigton
accreting black hole can wipe out some 10 per cent of the mass of an
optically thick spheroid simply by radiation pressure.  Also this
self-limiting mechanism induces a black-hole bulge relation compatible
with observations.  Repeating their calculation for our $10^{11}$
\msun\ case spheroid we obtain $f_s < 0.06 {\cal C}$.  Clearly, the
presence of an outward-moving shell with ${\cal C}=1$ makes the case
for a massive wind driven by radiation pressure much more convincing.

\section{Shining quasars in the hierarchical context}

All the calculations given above assume that an Eddington-accreting
black hole is present in a star-forming spheroid, and do not take into
proper account two very important aspects of galaxy and quasar
formation, namely the formation and evolution of galaxies driven by
the hierarchical assembly of dark matter halos, and the nearly
complete loss of angular momentum necessary to the gas to accrete onto
the central black hole.  We show in the following how the results of
Section 3 apply when inserted into a hierarchical galaxy formation
model, with simple assumptions on the loss of angular momentum.  

To show in which conditions quasar-triggered winds will lead to a
self-limited black hole--bulge relation, we present an example
obtained by inserting reasonable rules for quasar accretion into a
hierarchical galaxy formation model.  This model is still under
construction, and will be presented in full detail elsewhere.  The aim
here is not to present a proper complete modeling of galaxies and
quasars, but just to illustrate how the triggering of galaxy winds may
behave in a more realistic situation than a ``monolithic'' spheroid.

\subsection{The galaxy formation model}

The model follows loosely the ``semi-analytic'' models of Kauffmann et
al. (2000), Cole et al. (2000), Sommerville, Primack \& Faber (2000),
Menci et al. (2002), Hatton et al. (2003), Kang et al. (2004).  Here
we give a very quick description of the ingredients used.

(i) Merger histories of dark matter halos are obtained with the {\sc
pinocchio} tool (Monaco et al. 2002; Monaco, Theuns \& Taffoni 2002;
Taffoni, Monaco \& Theuns 2002).
(ii) The survival and merger of substructure (galaxies) is modeled
using the results of Taffoni et al. (2003).
(iii) Gas in the dark matter halos is shock-heated, and is assumed
to be in hydrostatic equilibrium.  It is treated as a polytropic gas
with polytropic index $1.1-1.3$.
(iv) Radiative cooling is described with the cooling function of
Sutherland \& Dopita (1993).  The evolution of the cooling radius is
followed taking into account also the hot gas emitted by
star-forming galaxies.
(v) The cooled gas is put into a cold halo gas component, that is then
let infall on the central galaxy in a few infall timescales.
(vi) Whenever the hot halo gas is hotter than the virial temperature,
it is ejected to the external space as a wind.  It is then acquired
back again by the halo when its virial temperature is equal to the
temperature of the gas at the escape time.
(vii) The same thing happens to the cold gas whenever its kinetic
energy overtakes the virial value.
(viii) The disc structure is computed with the Mo, Mao \& White (1998)
model, taking into account the presence of a bulge.
(ix) Star formation and feedback in discs are computed with the aid of
an extended Schmidt law, based on the results of paper I.  Feedback
gives thermal energy to the hot halo gas and kinetic energy to the
cold halo gas.
(x) Mergers bring gas and stars of the satellite galaxy to the bulge
of the central galaxy, in major mergers all the mass of the galaxies
is given to the bulge of the central galaxy.  Bulge radii are computed
following Cole et al. (2000).
(xi) Disc instabilities bring half of the disc mass into the bulge.
(xii) The evolution of the galaxy is followed by numerically
integrating a system of equations that follows the dynamics of all the
components.

A seed black hole of 1000 \msun\ is put in the centre of each dark
matter halo (see, e.g., Rees 2004 for a discussion on how seed black
holes are generated).  All the results depend very weakly on the
precise mass of the seed black hole, as long as it is not too small.

To be able to accrete onto a black hole, the gas must lose nearly all
of its angular momentum.  The first loss is connected to the
large-scale tidal fields responsible for the formation of a spheroid,
due to merger events or disc instabilities.  Some low-angular momentum
gas can also be available due to the specific distribution of angular
momentum within the dark matter halo (Kouschiappas, Bullock \& Dekel
2004); in this paper we will not take this possibility into account.
The origin of further losses for the cold gas residing in a bulge is
unclear.  Umemura (2001) proposed that the radiation drag from the UV
field of young stars can lead to the creation of a low-angular
momentum reservoir of gas, available for accretion onto the black
hole.  In this case, a fraction \kres\circa$10^{-3}$ of the star
formation rate can accrete onto the black hole.  The interesting point
is that many reasonable processes can lead to a linear relation
between the creation of the reservoir and the star formation rate (or,
equivalently, the amount of cold gas).  For instance, turbulence or
magnetic fields may be responsible for the loss of angular momentum,
and these events are again associated to the formation of massive
stars and their consequent explosion as SNe.

If the accretion rate induced by star formation is larger than the
Eddington limit, the exceeding gas is put into a reservoir, that is
later accreted onto an Eddington time:

\be \Dmbh  = \min\left(\Kres\Dmsfr +\frac{M_{\rm resv}}{t_{\rm ed}}\, ,
\, \frac{M_\bullet}{t_{\rm ed}} \right)\, , \label{eq:dmbh} \ee
\be \dot{M}_{\rm res} = \Kres\Dmsfr - \Dmbh\, . \label{eq:dmres} \ee

\noindent
At each merger the black holes are immediately merged as well, and at
major mergers the reservoir is destroyed.  Granato et al. (2004) have
presented a more refined modeling of this reservoir; we prefer however
keep a simpler approach here.

The modeling of star formation in bulges is determinant to assess
whether quasars are able to trigger galaxy winds.  Most semi-analytic
codes assume very quick star-formation timescales, of order of the
bulge dynamical time.  We have shown in Section 3 that following a
merger, while it is reasonable that a significant amount of gas is
quickly consumed into stars, star-formation timescales will be rather
long as soon as self-regulation is achieved.  A reasonable compromise
is given by forcing the starburst to follow the observational Schmidt
law (Schmidt 1959; Kennicutt 1998).  The model presented in paper I
indeed follows roughly the Schmidt-law, although with a lower
normalization (figure~7 of paper I shows that star formation is below
the Kennicutt relation if the effective vertical scalelength, here
identified with the half-mass radius of the bulge, is of order of 1
kpc).  In this case the star-formation timescale is:

\be \tau_\star = 2.2\times 10^{8} 
\frac{M_{\rm cold}}{10^{11}\ {\rm M}_\odot}
\left( \frac{R_{\rm hm}}{4.9\ {\rm kpc}} \right)^{-2}
\ {\rm yr} \, . \label{eq:taust} \ee

\begin{figure*}
\centerline{
\includegraphics[width=8.4cm]{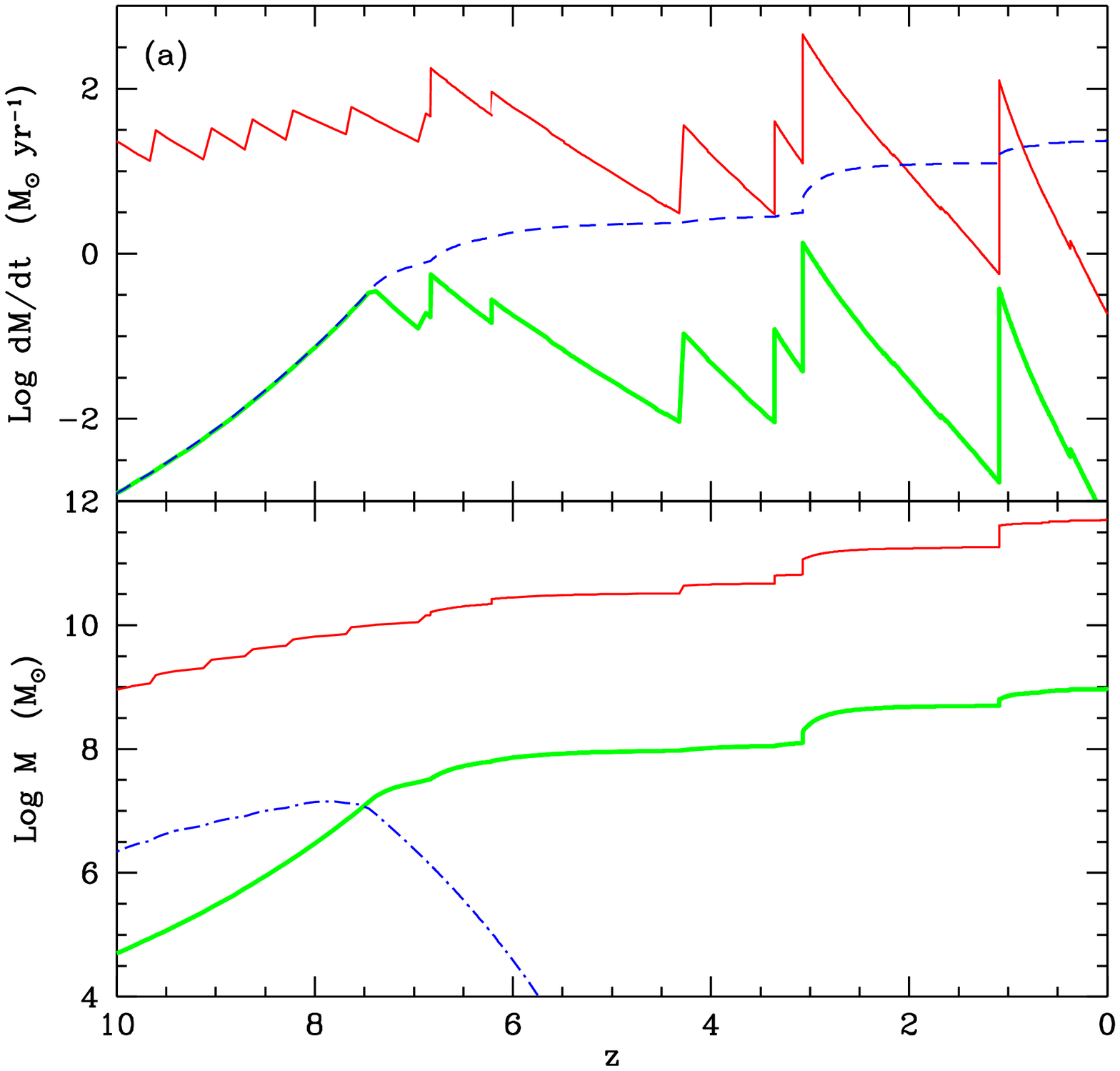}
\includegraphics[width=8.4cm]{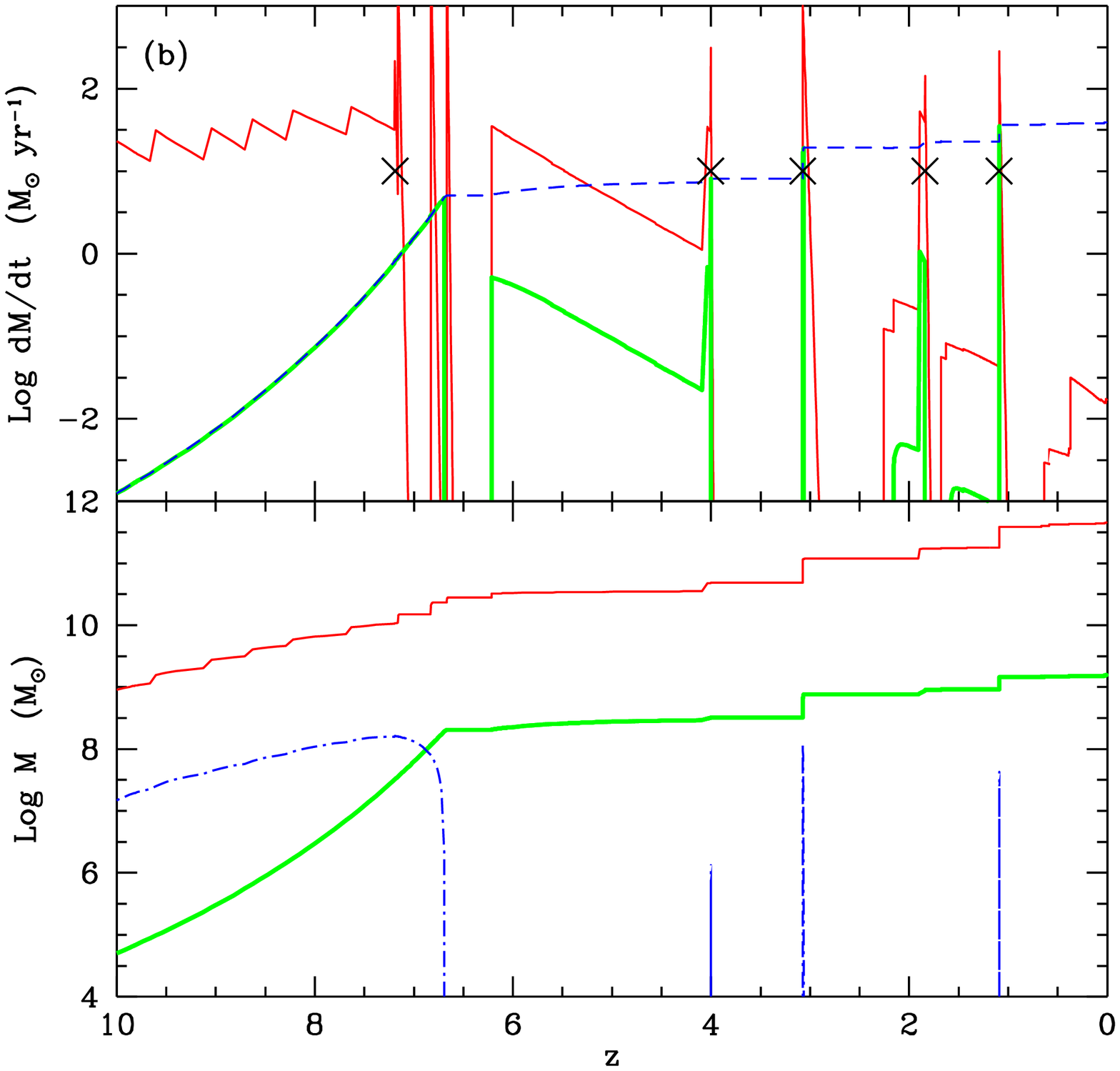}}
\caption{Evolution of the main progenitor of an elliptical galaxy
contained in a $1.1\times 10^{13}$ \msun\ dark matter halo.  (a): no
winds (\kres=0.003).  (b): with winds (\krh=50, \kres=0.02 and
\ktr=1).  Upper panels: star formation rate in the bulge of the main
progenitor (continuous lines), accretion rate on the black hole (heavy
continuous lines), Eddington limit (dashed lines).  The crosses in the
right panel denote the shining events.  Lower panels: stellar mass of
the bulge of the main progenitor (continuous lines), black hole mass
(heavy continuous lines), reservoir mass (dot-dashed lines).}
\end{figure*}

\subsection{Low-angular momentum gas and the triggering of the wind}

As long as a significant fraction of the stars of a spheroid is formed
within the spheroid itself (and not in self-regulated discs that later
merge into spheroids), the relation between star formation rate and
black hole accretion (equation~\ref{eq:dmbh}) induces a black
hole--bulge correlation similar to that observed for reasonable values
of \kres.  Many authors (cited in the introduction) have used similar
rules in conjunction with hierarchical galaxy formation models,
correctly reproducing the black hole--bulge correlation with no need
for self-regulation.  With our galaxy formation model we find a good
fit for the black hole--bulge correlation for \kres=0.003.  We
demonstrate now that in this case galaxy winds are likely irrelevant.

According to Figure~2, the catastrophic switch from the adiabatic to
the PDS confinement regime happens whenever the evaporation rate
overtakes the (unperturbed) star-formation rate by a factor of
\circa10.  However, this is valid in the region affected by radiative
heating, which for simplicity we assumed in Section 3 to coincide with
the whole galaxy; in a realistic case the evaporation is limited to
the inner region of the galaxy, $r<R_{\rm rh}$, which contains only a
fraction of the total cold gas and star formation rate (\circa20 per
cent in our $10^{11}$ \msun\ spheroid case).  Even within this region,
the evaporation grows linearly with the radius $r$ through the
covering factor ${\cal C}$ (equation~\ref{eq:krh}), so if the gas
distribution in this region is flatter than $r^{-2}$ (and the mass of
cold gas grows more rapidly than $r$) the triggering condition will
first be reached in the inner regions and then propagate
inside-out\footnote{Besides, the condition of adiabatic confinement
requires that the gas distribution is flatter than $r^{-2}$ up to the
vertical scale-length of the system.}.  A reasonable and simple
triggering criterion for a star-forming galaxy is then obtained when
evaporation overtakes the {\it global} star-formation rate by a factor
\ktr\ of order unity:

\be \Dmrh > \Ktr\Dmsfr\, . \label{eq:trigger} \ee

Using equations~\ref{eq:rh}, \ref{eq:dmbh} (assuming that accretion is
not Eddington limited) and \ref{eq:trigger}, it is easy to see that
winds will be triggered if $\Ktr<\Krh\times\Kres$.  For \kres=0.003
(the value with which the black hole--bulge correlation is reproduced
with no winds) and \krh=50, this implies $\Ktr<0.15$.  In other words,
winds will work only if radiative heating is concentrated on \circa1.5
per cent of the star-forming region.  The result of switching winds in
this case is to limit black hole masses to lower values, with a
resulting underestimate.

Alternatively, the mass deposition rate on the reservoir can be
higher.  In this case the black hole--bulge correlation is due to the
self-limiting action of winds.  Keeping \krh=50 and assuming \ktr=1,
winds are triggered if $\Kres\ge 0.02$, nearly an order of magnitude
higher than before.

To be more specific, a massive wind in the galaxy formation model is
triggered whenever: (i) the triggering condition of
equation~\ref{eq:trigger} is satisfied; (ii) the accretion rate in
Eddington units is larger than 0.01 (otherwise the system switches to
a radiatively inefficient accretion mode, and the present model does
not apply); (iii) the amount of cold mass present in the bulge is
lower than equation~\ref{eq:f_s}.  To implement the post-wind scenario
described in Section 3, 10 per cent of the cold gas is not put into
the shell but consumed by star formation on a bulge dynamical time,
while a fraction \kres\ of this gas is put in the reservoir to be
accreted onto the black hole.  
The cold gas in the shell is then given to the cold halo component.

The main uncertainties and degrees of freedom in the modeling of the
wind are the following: (i) the timescale of star formation in bulges
is fixed in a phenomenological way, better modeling is needed.  (ii)
The modeling of the reservoir of low-angular momentum gas is
reasonable but not unique.  (iii) The parameter \krh\ is fixed with no
explicit reference to $R_{\rm rh}$ or to the covering factor of cold
clouds.  (iv) The parameter \kres\ is very poorly constrained.  (v)
The constant in equation~\ref{eq:f_s} depends on many uncertain
details and may reasonably be considered as a free parameter (we will
keep it fixed to the value given above).  (vi) The parameter \ktr\ is
uncertain by at least a factor of ten.  (vii) The fraction of mass
compressed to the centre after the wind is uncertain as well.

\begin{figure}
\centerline{
\includegraphics[width=8.4cm]{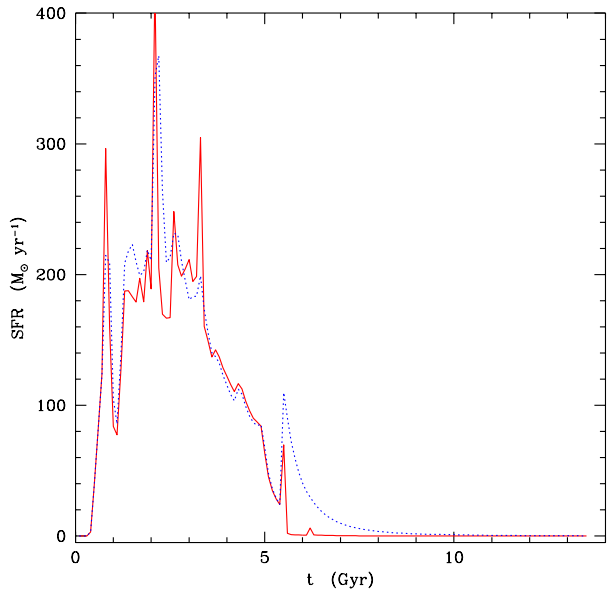}}
\caption{Star formation rates for the two examples shown in Figure 3.
This time the star formation rate of all the stars contained in the
galaxy at $z=0$ is given.  The dotted line refers to the example
without winds (Figure 3a), the continuous line to the
example with winds (Figure 3b).}
\label{fig:duesfr}
\end{figure}

\begin{figure}
\centerline{
\includegraphics[width=8.4cm]{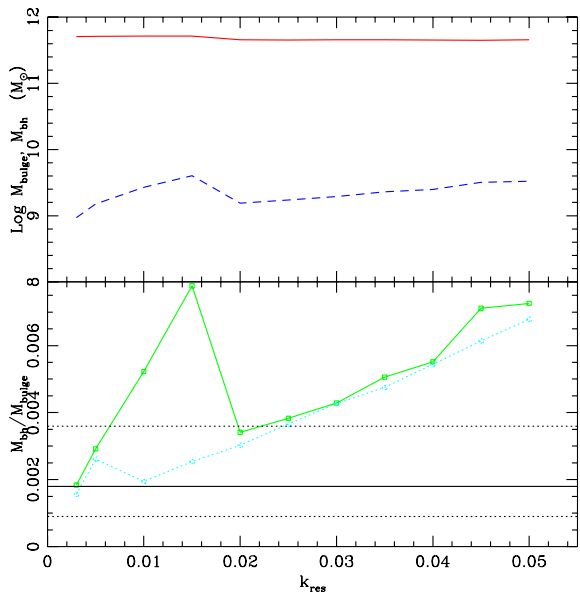}}
\caption{Upper panel: bulge (continuous line) and black hole (dashed
line) masses as a function of \kres\ for the example of Figure 3
(\ktr=1).  Lower panel: black hole--bulge ratio as a function of
\kres\ for the case \ktr=1 (continuous line) and \ktr=0.3 (dotted
line).  The horizontal continuous and dashed lines show the range
allowed by observations (Shankar et al. 2004).}
\label{fig:stab}
\end{figure}

\subsection{An example}

In this subsection we show that there is a significant part of the
parameter space that allows for quasar-triggered galaxy winds, and
briefly outline the effect these would have on the galaxy population
as a whole.

We have taken a $1.1\times 10^{13}$ \msun\ dark matter halo from a
$256^3$ {\sc pinocchio} realization of size 93 Mpc.  The particle mass
for this box is $3\times 10^9$, corresponding to a mass resolution for
the merger tree of \circa$10^{11}$ \msun.  Due to its particular
merger history, this halo hosts a spheroid at the final time ($z=0$).
The mass of the spheroid depends of course on the details of galaxy
formation, and ranges from $10^{11}$ to $5\times10^{11}$ \msun.  We
have run the model assuming first no winds (\krh=0), so that the black
hole--bulge correlation is set by angular momentum loss
(\kres=0.003).   Figure~3a shows the
resulting evolution of the main progenitor of the galaxy.  In
particular, we show accretion and star formation rates, bulge, black
hole and reservoir masses.  It is worth clarifying that all these
quantities refer to the bulge of the main progenitor; accretion takes
places also in the black holes of satellite galaxies, while stars form
both in satellites and discs.

In this case of no wind, the assembly of the galaxy proceeds
gradually.  The two main star-formation events are located at
$z$\circa3 and \circa1, while the sawtooth-like events at very high
redshift are due to successive disc instabilities.  The black hole
starts accreting mass at very early times, and its accretion is
Eddington limited till $z$\circa7.5.  By $z$\circa6 a \circa$10^8$
\msun\ black hole is already present.  The two main shining events
closely follow the mergers, and are both sub-Eddington.  Most black
hole mass is acquired by accretion more than by black hole mergers.
At the final time the bulge and black hole masses are $5.1\times
10^{11}$ and $9.4\times10^8$ \msun, compatible with the observational
black hole--bulge relation.

Figure~3b shows the same example with \krh=50 and \kres=0.02.  The
evolution at very high redshift is similar, but the Eddington
accreting phase goes on for a longer time (due to the higher amount of
low-angular momentum gas available), so that the resulting black hole
is a factor of 3 more massive at redshift \circa6.  The first shining
event takes place at $z$\circa7, and has the effect of quenching star
formation in the spheroid.  Further shinings take place at redshifts
\circa4, 3, 2 and 1.  In each case we have a short quasar phase that
however contributes significantly at the mass of the black hole only
at redshift 3 (with a significant burst at $z$\circa 1).  In the
meantime, the episodes of star formation are quickly quenched.  As a
result, the final bulge and black hole masses are $4.6\times10^{11}$
and $1.6\times10^9$ \msun.  This galaxy lies a factor of 2 above the
observational relation, just below the 1-$\sigma$ 0.3 dex dispersion.

Figure~4 shows, as a function of cosmological time, the star formation
rates of all the stars contained in the final spheroid for the two
cases.  This is different from the star formation rate of the bulge of
the main progenitor, as it contains the contribution of stars formed
in discs and satellites.  The two star formation rates do not differ
much; the most notable effect is the truncation of the low-redshift
tail of star formation.  This detail however can be very important
because minor episodes of star formation can influence strongly the
luminosity-weighted colours of galaxies.  Moreover, the chemical
evolution of these object is also affected by late star formation
events, that take place from iron-enriched material and decrease the
$\alpha$-enhancement (see, e.g., Matteucci 1996).  The effect of winds
will then be to increase the number of passive old galaxies at
$z$\circa1--2 and to allow a high level of $\alpha$-enhancement.

Besides, the quasar population will be more deeply affected by winds,
especially for $M_{\rm bul}<10^{10}$ \msun, where SNe alone are able
to generate a massive wind.  In this case we expect an increase of
bright quasars at very high redshift (because the Eddington limited
phase lasts longer) and a decrease at low redshift (because the
shining events are suppressed by winds), in agreement with the trends
suggested by observations.

Figure~5 shows how the black hole and bulge masses and their ratio
change when the \kres\ parameter is increased; for \kres$<$0.02 both
black hole mass and ratio grow linearly with \kres, but as soon as
winds come into play the fraction self-regulates to a value that
remains relatively stable till \kres=0.03, then starts growing again.
The upper panel shows that the main effect of wind triggering is on
the black hole, though bulges tend to be slightly less massive.
Finally, a good value for the black hole--bulge ratio is obtained with
\ktr=0.3 and \kres=0.01 (dotted line)

\section{Conclusions}

Radiation pressure and radiative heating are two unavoidable processes
during the shining of a quasar that strongly influence the ISM of a
star-forming spheroid.  However, taken alone they may not be able to
trigger those massive winds that have often been assumed to take place
during the formation of a spheroidal galaxy.  Based on a model for
feedback in galaxy formation, we have proposed that a massive removal
of ISM from a large star-forming spheroid can be triggered by the
joint action of SNe and quasar light as follows: (i) stars form in a
self-regulated way in a two-phase, highly pressurized ISM; (ii)
through runaway radiative heating, the quasar evaporates part of the
cold phase; (iii) when the evaporation rate is strong enough, the
feedback regime changes: due to the higher density of the hot phase,
SBs arising from the star-forming clouds get to the PDS stage before
being halted by pressure confinement; (iv) the consequent drop in
pressure leads to the percolation of cold shells and to the creation
of an expanding super-shell; (v) some mass is compressed to the
centre, giving rise to a nuclear starburst and to further black hole
accretion; (vi) the remaining cold gas (not included in the shell or
in the nucleus) is most likely involved in diffuse star formation;
(vii) radiation pressure pushes the shell out of the galaxy if it is
not too massive; (viii) as soon as it interacts with the external hot
halo gas, the shell halts and fragments; (ix) a part of it is finally
evaporated by radiative heating, 
though to a much lower temperature than the inverse Compton one of the
quasar.

We have demonstrated that this mechanism can lead to a self-limited
black hole--bulge relation similar to the observed one.  However,
using a galaxy formation model (presently under development) that
takes into account black hole accretion, we have shown that the black
hole--bulge relation is also reproduced assuming that the rate of
deposition of low-angular momentum gas onto the black hole amounts 0.3
per cent of the star formation rate in the bulge.  Including a
motivated criterion for quasar-triggered winds, we have shown that the
black hole--bulge relation is self-limited by winds whenever the
deposition rate is at least \circa1 per cent of the bulge
star-formation rate.  Compared to the no-wind case, this mechanism
leads to more massive black holes at high redshift and to a quenching
of low-redshift activity.

There are two main possible objection to this scenario.  First, a
massive removal of ISM can be caused by other mechanisms, like a
kinetic ejection of matter from the AGN (Granato et al. 2004) or
simply by radiative heating (Sazonov et al. 2004a) or radiation
pressure (Murray et al. 2005) by the quasar light, so there is no real
need for such a sophisticated and indirect mechanism.  Ejection of
matter at nearly relativistic speed is observed in action in extreme
BAL quasars; however, this mechanism can work only if the covering
angle of the outflow is large.  On the other hand, the radiation of
the quasar alone may be able to remove only a modest fraction of the
mass of the ISM, not larger than a few per cent.  The triggering
mechanism suggested here has the merit of creating an outwardly
expanding, optically thick super-shell (with unity covering factor),
which is then easily pushed away by radiation pressure.  An
appreciable element is that the prediction comes out naturally from
the model of paper I without any parameter tuning.

The second objection is that if a reservoir of low angular momentum
gas is created at a rate proportional to the star-formation rate in
bulges, then there is no need for a self-limiting mechanism
responsible for the black hole--bulge relation.  In this case the
condition for wind triggering may never (or almost never) be reached
in practice.  Probably the strongest argument in favour of winds lies
in the chemical enrichment patterns of bulges (stars, ISM and
circum-quasar gas), but more work is needed for an assessment of this
point.

It is not easy to device critical observational tests to understand if
a mechanism like the one described here is actual.  Due to the number
of processes involved and to the huge uncertainties in many of the
parameters, many different configurations may lead to very similar
predictions.  To be more specific, all the evolution from the
starburst to the ejection of the shell would be hidden by dust, so the
quasar would be invisible.  Shining quasars would correspond to the
stage after the destruction of the shell, so the only clearly
observable stage would correspond to the last phases of expulsion and
evaporation of the shell.  For \krh=50 this takes a time $t\sim
5\times 10^8 f_s M_{\rm bul,11} \dot{M}_{\bullet,4}$ yr, so a massive
shell would be visible for \circa$12\times f_s$ Eddington times as an
absorber with a low expansion velocity.  Such objects are seen for
instance in absorption of optical quasar spectra (see, e.g., Srianand
\& Petitjean 2000; D'odorico et al. 2004).  Their structure is
generally very complex, and outflowing speeds range from hundreds to
tens of thousands of km s$^{-1}$.  A quick and dirty comparison with
such data is then impossible, and more precise predictions need a
dedicated calcolation and will be presented elsewhere.

Another way to probe the validity of the wind model is through its
effects on the statistical properties of galaxies.  In this case the
problem is in the degeneracy with all the other (numerous and
uncertain) parameters of galaxy formation.  Anyway, the higher
accretion rate connected with quasar triggered winds gives more
massive black holes at $z\sim6$ and a quenching of low-$z$ activity,
in line with the observational evidence, while the expulsion of the
metals generated after the first burst of star formation is in line
both with the high level of alpha enhancement of ellipticals and with
the high abundance of iron in clusters.  Moreover, the fact that no
limit on the fraction of mass that can be removed is present for
bulges less massive than $10^{10}$ \msun\ implies an increase in the
scatter of the black hole--bulge relation at small masses; the
observations for such small bulges are few, but the possible finding
of outliers in the relation would be in line with this prediction.  A
more accurate analysis will be presented in a subsequent paper.

\section*{Acknowledgments}
We thank Giuliano Taffoni for his help in developing the galaxy
formation code, and Stefano Cristiani, Mitch Begelman and Gianluigi
Granato for discussions.

{}

\bsp

\label{lastpage}


\begin{thebibliography}{}
\bibitem[]{ab} Begelman M.C., 1985, ApJ, 297, 492
\bibitem[Begelman(2004)]{2004cbhg.symp..375B} Begelman M.C., 2004, in L.C. Ho ed., Coevolution of Black Holes and Galaxies. Cambridge University Press, Cambridge, p. 375
\bibitem[]{ac} Begelman M.C., McKee C.F., Shields G.A., 1983, ApJ, 271, 70
\bibitem[]{ad} Bromley J.M., Somerville R.S., Fabian A.C., 2004, MNRAS, 350, 456
\bibitem[]{ae} Cattaneo A., 2001, MNRAS, 324, 128
\bibitem[]{af} Cavaliere A., Vittorini V., 2002, ApJ, 570, 114 
\bibitem[]{ai} Cioffi, D.F., McKee, C.F., Bertschinger, E., 1988, ApJ, 334, 252
\bibitem[Ciotti \& Ostriker(1997)]{1997ApJ...487L.105C} Ciotti L., Ostriker J.P., 1997, ApJ, 487, L105
\bibitem[]{ia} Cole S., Lacey C.G., Baugh C.M., Frenk C.S., 2000, MNRAS, 319, 168
\bibitem[]{al} Comastri A., Setti G., Zamorani G., Hasinger G., 1995, A\&A, 296, 1
\bibitem[]{ah} Cristiani S., Vio R., 1990, A\&A, 227, 385
\bibitem[]{do} D'odorico V., Cristiani S., Romano D., Granato G.L., Danese L., 2004, MNRAS, 351, 976
\bibitem[Dunlop et al.(2003)]{2003MNRAS.340.1095D} Dunlop J.S., McLure R.J., Kukula M.J., Baum S.A., O'Dea C.P., Hughes D.H., 2003, MNRAS, 340, 1095 
\bibitem[Edge \& Frayer(2003)]{2003ApJ...594L..13E} Edge A.C., Frayer D.T., 2003, ApJ, 594, L13 
\bibitem[]{an} Elmegreen B.G., 2002, ApJ, 564, 773
\bibitem[]{ao} Fabian A.C., 1999, MNRAS, 308, 39
\bibitem[]{ap} Granato G.L., De Zotti G., Silva L., Bressan A., Danese L., 2004, ApJ, 600, 580
\bibitem[]{aq} Granato G.L., Silva L., Monaco P., Panuzzo P., Salucci P., De Zotti G., Danese L., 2001, MNRAS, 324, 757
\bibitem[]{ar} Haehnelt M.G., Natarajan P., Rees M.J., 1998, MNRAS, 300, 817
\bibitem[Haiman et al.(2004)]{2004ApJ...606..763H} Haiman Z., Ciotti L.,  Ostriker J.P., 2004, ApJ, 606, 763 
\bibitem[Hamann \& Ferland(1999)]{1999ARA&A..37..487H} Hamann F., Ferland G., 1999, ARA\&A, 37, 487 
\bibitem[]{as} H\"aring N., Rix H.W., 2004, ApJ, 604, 89
\bibitem[]{at} Hatton S., Devriendt J.E.G., Ninin S., Bouchet F.R., Guiderdoni B., Vibert D., 2003, MNRAS, 343, 75
\bibitem[Hatziminaoglou et al.(2003)]{2003MNRAS.343..692H} Hatziminaoglou E., Mathez G., Solanes J., Manrique A., Salvador-Sol{\' e} E., 2003, MNRAS, 343, 692 
\bibitem[]{au} Kang X., Jing Y.P., Mo H.J., Boerner G., 2004, astro-ph/0408475
\bibitem[]{av} Kauffmann G., Haehnelt M.G., 2000, MNRAS, 311, 576
\bibitem[]{az} Kauffmann G., Colberg J.M., Diaferio A., White S.D.M., 1999, MNRAS, 307, 529
\bibitem[]{bd} Kennicutt R.C., 1998, ApJ, 498, 541
\bibitem[]{be} Kormendy J., Richstone D., 1995, ARA\&A, 33, 581
\bibitem[Koushiappas, Bullock, Dekel(2004)]{2004MNRAS.tmp..349K} Koushiappas S.M., Bullock J.S., Dekel A., 2004, MNRAS, 349 
\bibitem[]{bf} Kriss G.A., Davidsen A., Zheng W., Lee G., 1999, ApJ, 527, 683
\bibitem[]{bg} Krolik J.H., McKee C.F., Tarter C.B., 1981, ApJ, 249, 422
\bibitem[]{bh} Lombardi M., Bertin G., 2001, A\&A, 375, 1091
\bibitem[]{bi} Magorrian J., Tremaine S., Richstone D. et al., 1998, ApJ, 115, 2285
\bibitem[]{bj} Mahmood A., Devriend J.E.G., Silk J., astro-ph/0401003
\bibitem[Maiolino et al.(2004)]{2004A&A...420..889M} Maiolino R., Oliva E., Ghinassi F., Pedani M., Mannucci F., Mujica R., Juarez Y., 2004, A\&A, 420, 889 
\bibitem[]{bk} Marconi A., Hunt L.K., 2003, ApJ, 589, 21
\bibitem[]{bm} Matteucci F., 1994, A\&A, 288, 57
\bibitem[Matteucci(1996)]{1996FCPh...17..283M} Matteucci F., 1996, Fund.  Cosm. Phys., 17, 283 
\bibitem[Matteucci \& Padovani(1993)]{1993ApJ...419..485M} Matteucci F., Padovani P., 1993, ApJ, 419, 485 
\bibitem[]{bo} Menci N., Cavaliere A., Fontana A., Giallongo E., Poli F., Vittorini V., 2003, ApJ, 587, 63
\bibitem[]{bp} Mo H.J., Mao S., White S.D.M., 1998, MNRAS, 295, 319
\bibitem[]{bq} Monaco P., Theuns T., Taffoni G., Governato F., Quinn T., Stadel J., 2002, ApJ, 564, 8
\bibitem[]{br} Monaco P., Theuns T., Taffoni G., 2002, MNRAS, 331, 587
\bibitem[]{bs} Monaco P., 2004a, MNRAS, 352, 181 (paper I)
\bibitem[]{bt} Monaco P., 2004b, MNRAS, 354, 151
\bibitem[]{mf} Monaco P., 2004c, in eds. A. Merloni, S. Nayakshin, R. Sunyaev, Growing Black Holes.   Springer-Verlag, Berlin, in press
\bibitem[]{bu} Mori M., Ferrara A., Madau P., 2002, ApJ, 571, 40
\bibitem[Murray et al.(2005)]{2005ApJ...618..569M} Murray N., Quataert E., Thompson T.A., 2005, ApJ, 618, 569
\bibitem[]{bz} Ostriker J.P., McKee C.F., 1988, RvMP,60, 1O
\bibitem[]{ca} Ostriker J.P., Ciotti L., 2004 (astro-ph/0407234)
\bibitem[]{op} Rees M., 2004, in eds. A. Merloni, S. Nayakshin, R. Sunyaev, Growing Black Holes.   Springer-Verlag, Berlin, in press
\bibitem[]{rr} Renzini A., 2004, in eds. J.S. Mulchaey, A. Dressler,  A. Oemler 2004, Clusters of Galaxies: Probes of Cosmological Structure and Galaxy Evolution. Carnegie Observatories Astrophysics Series.  Pag 261 
\bibitem[]{cb} Risaliti G., Elvis M., 2004, (astro-ph/0403361)
\bibitem[]{cc} Romano D., Silva L., Matteucci F., Danese L., 2002, MNRAS, 334, 444
\bibitem[]{cd} Salucci P., Szuszkiewicz E., Monaco P., Danese L., 1999, MNRAS, 307, 637
\bibitem[]{yu} Sazonov S.Yu., Ostriker J.P., Ciotti L., Sunyaev R.A., 2004a, submitted to MNRAS (astro-ph/0411086)
\bibitem[]{ce} Sazonov S.Yu., Ostriker J.P., Sunyaev R.A., 2004b, MNRAS, 347, 144
\bibitem[]{cf} Schmidt M., 1959, ApJ, 129, 243
\bibitem[]{ch} Silk J., Rees M.J., 1998, A\&A, 331, 1
\bibitem[]{ci} Shankar F., Salucci P., Granato G.L., De Zotti G., Danese L., 2004, MNRAS, 354, 1020
\bibitem[]{cj} Somerville R.S., Primack J.R., Faber S.M., 2001, MNRAS, 320, 504
\bibitem[]{sh} Springel V., Di Matteo T., Hernquist L., 2004, submitted to MNRAS (astro-ph/0411108)
\bibitem[]{sr} Srianand R., Petitjean P., 2000, A\&A, 357, 414
\bibitem[]{ck} Sutherland R.S., Dopita M.A., 1993, ApJS, 88, 253
\bibitem[]{cl} Taffoni G., Mayer L., Colpi M., Governato F., 2003, MNRAS, 341, 434
\bibitem[]{cm} Taffoni G., Monaco P., Theuns T., 2002, MNRAS, 333, 623
\bibitem[]{co} Thomas D., 1999, MNRAS, 306, 655
\bibitem[]{cp} Umemura M., 2001, ApJ, 560, 29
\bibitem[]{cq} Vignali C., Brandt W.N., Schneider D.P., 2003, ApJ, 125, 433
\bibitem[]{cr} Yu Q., Tremaine S., 2002, MNRAS, 335, 965
\bibitem[]{cs} Weaver R., McCray R., Castor J., Shapiro P., Moore R., 1977, ApJ, 218, 377


\end{thebibliography}
\end{document}